\documentclass[twocolumn,prb]{revtex4}

\usepackage{graphicx}
\usepackage{dcolumn}
\usepackage{bm}

\begin{document}


\title{Experimental Realization of Spin-$\frac{\bf 1}{\bf 2}$ Triangular-Lattice Heisenberg Antiferromagnet}

\author{Yutaka Shirata$^1$}
\author{Hidekazu Tanaka$^1$} 
\author{Akira Matsuo$^2$}
\author{Koichi Kindo$^2$}

\affiliation{
$^1$Department of Physics, Tokyo Institute of Technology, Meguro-ku, Tokyo 152-8551, Japan\\
$^2$Institute for Solid State Physics, University of Tokyo, Kashiwa, Chiba 277-8581, Japan\\
}


\begin{abstract}
We report the results of magnetization and specific heat measurements on Ba$_3$CoSb$_2$O$_9$, in which the magnetic Co$^{2+}$ ion has a fictitious spin-$\frac{1}{2}$, and show evidence that a spin-$\frac{1}{2}$ Heisenberg antiferromagnet on a regular triangular lattice is actually realized in Ba$_3$CoSb$_2$O$_9$. We found that the entire magnetization curve including the one-third quantum magnetization plateau is in excellent agreement with theoretical calculations at a quantitative level.
\end{abstract}

\maketitle


Exploring the ground state of a frustrated quantum magnet has been one of the main subjects of condensed matter physics~\cite{Anderson,Kalmeyer,Balents}. A long theoretical debate reached a consensus that a two-dimensional (2D) spin-$\frac{1}{2}$ TLHAF has an ordered ground state of the 120$^{\circ}$ spin structure with an extremely small sublattice magnetization~\cite{Huse,Jolicoeur,Bernu,Singh}. Although the zero-field ground state is qualitatively the same as that for the classical spin, the ground state in a magnetic field ${\bm H}$ cannot be determined uniquely only from the classical model. The classical equilibrium condition is given by ${\bm S}_1\,{+}\,{\bm S}_2\,{+}\,{\bm S}_3\,{=}\,g{\mu}_{\rm B}{\bm H}/(3J)$ with the sublattice spin Si. Because there are an infinite number of spin states that satisfy this condition, the classical ground state is continuously degenerate. The ground state of a small spin TLHAF in a magnetic field is essentially determined by the quantum fluctuation energy. A remarkable quantum effect is that an up-up-down spin state, which appears in a magnetic field for the classical model, can be stabilized in a finite magnetic field range, so that the magnetization curve has a plateau at one-third of the saturation magnetization~\cite{Nishimori,Chubukov,Nikuni,Alicea,Farnell,Honecker,Sakai}.

The nature of the quantum mechanical ground state in a magnetic field is strongly reflected in the magnetization process. The magnetization process for a 2D spin-$\frac{1}{2}$ TLHAF, which exhibits the most pronounced quantum effect, was calculated energetically by means of spin wave theory~\cite{Chubukov,Alicea}, coupled cluster method~\cite{Farnell} and exact diagonalization~\cite{Honecker,Sakai}. The calculated magnetization curves are greatly different from that for the classical spin. However, experimental verification of the theoretical results has not been conducted at a quantitative level.

Experimentally, Cs$_2$CuCl$_4$~\cite{Coldea}, Cs$_2$CuBr$_4$~\cite{Ono1,Fortune} and $\kappa$-(BEDT-TTF)$_2$Cu$_2$(CN)$_3$~\cite{Shimizu} have been actively investigated as spin-$\frac{1}{2}$ TLHAFs. However, the triangular lattice in these substances is not regular but distorted, and thus, the exchange interaction is spatially anisotropic. Cs$_2$CuCl$_4$ and Cs$_2$CuBr$_4$ also exhibit a large antisymmetric interaction of the Dzyaloshinsky-Moriya (DM) type. Although the quantum magnetization plateau has been actually observed in Cs$_2$CuBr$_4$~\cite{Ono1,Fortune}, the magnetization process is anisotropic and the magnetization plateau is not observed for a magnetic field perpendicular to the triangular lattice plane. In this letter, we present the results of magnetization and specific heat measurements on Ba$_3$CoSb$_2$O$_9$ and provide evidence that Ba$_3$CoSb$_2$O$_9$ closely approximates to the ideal spin-$\frac{1}{2}$ TLHAF.

\begin{figure}[htbp]
\begin{center}
\vspace{-2 cm}
\includegraphics[scale =0.40]{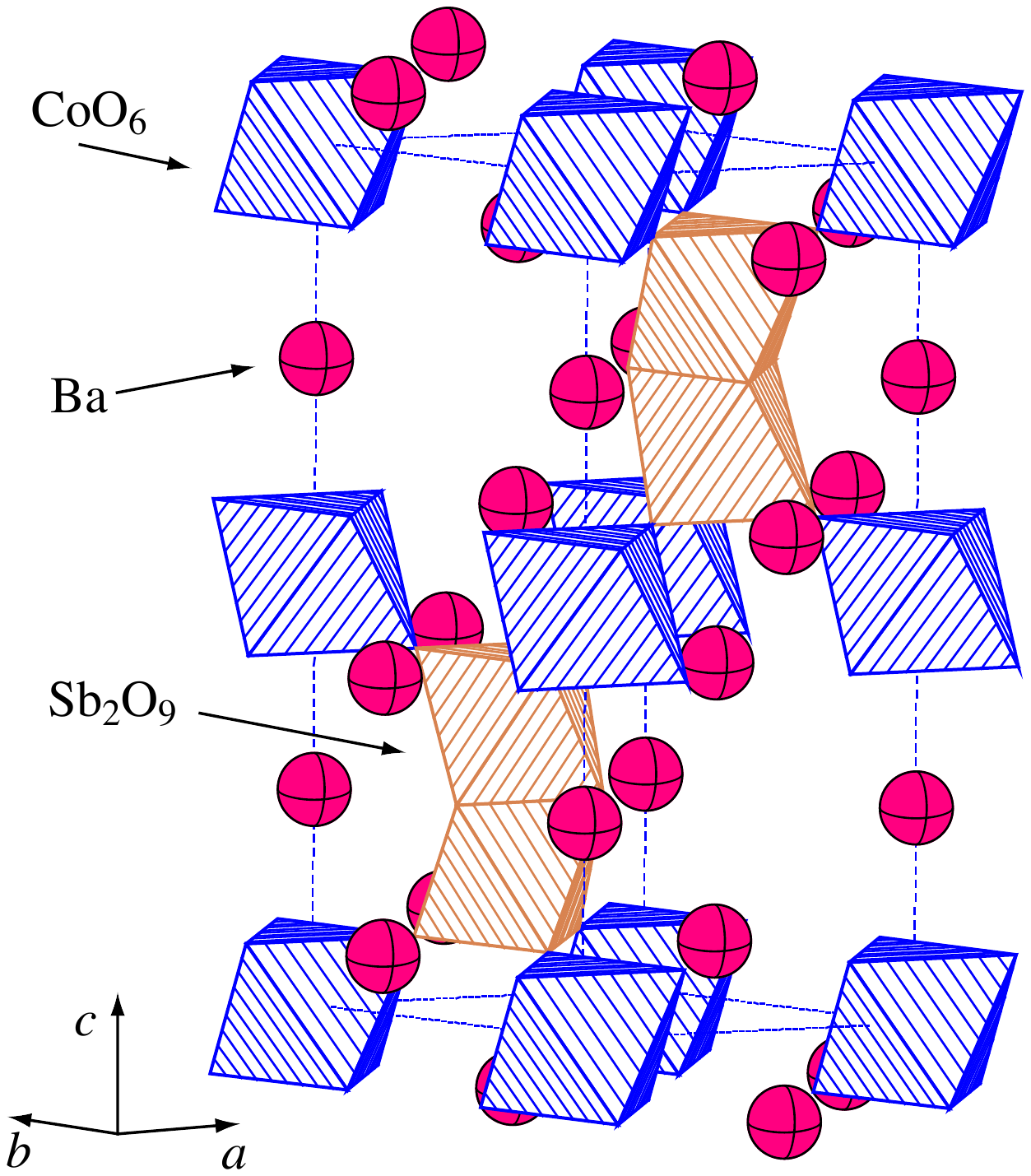}
\end{center}
\vspace{-4 cm}
\caption{(Color online) Crystal structure of Ba$_3$CoSb$_2$O$_9$. The blue single octahedron is a CoO$_6$ octahedron centred by Co$^{2+}$ and the face-sharing Sb$_2$O$_9$ double octahedron is shaded ochre. Magnetic Co$^{2+}$ ions form a regular triangular lattice in the ab plane. Dotted lines denote the chemical unit cell.}
 \label{fig:cryst}
\end{figure} 

Figure 1 shows the crystal structure of Ba$_3$CoSb$_2$O$_9$. This substance crystallizes in a highly symmetric hexagonal structure, $P6_3/mmc$, which is closely related to the hexagonal BaTiO$_3$ structure~\cite{Treiber,Doi}. The structure is composed of a single CoO$_6$ octahedron and a face-sharing Sb$_2$O$_9$ double octahedron, which are shaded blue and ochre, respectively. Magnetic Co$^{2+}$ ions form regular triangular lattice layers parallel to the ab plane, which are separated by the nonmagnetic layer of the Sb$_2$O$_9$ double octahedron and Ba$^{2+}$ ions. Therefore, the interlayer exchange interaction is expected to be much smaller than the intralayer exchange interaction. However, Ba$_3$CoSb$_2$O$_9$ undergoes magnetic ordering at $T_{\rm N}\,{\simeq}\,3.8$ K owing to the weak interlayer interaction~\cite{Doi}. Because of the high symmetric crystal structure, the antisymmetric DM interaction is absent between the first-, second- and third-neighbor spins in the triangular lattice and between all spin pairs along the $c$ axis.

It is known that the magnetic property of Co$^{2+}$ in an octahedral environment is determined by the lowest orbital triplet $^4T_1~$\cite{Abragam,Lines,Shiba}. This orbital triplet splits into six Kramers doublets owing to the spin-orbit coupling and the uniaxial crystal field, which are expressed together as
\begin{eqnarray}
{\cal H}^{\prime}=-(3/2){\lambda}({\bm l}\cdot{\bm S})-{\delta}\left\{(l^z)^2-2/3\right\},
\label{eq:perturb}
\end{eqnarray}
where $\bm l$ is the effective angular momentum with $l=1$ and $\bm S$ is the true spin with $S=\frac{3}{2}$. When the temperature $T$ is much lower than the magnitude of the spin-orbit coupling constant ${\lambda}\,{=}\,{-}\,178$ cm$^{-1}$, i.e., $T\,{\gg}\, |{\lambda}|/k_{\rm B}\,{\simeq}\,250$\,K, the magnetic property is determined by the lowest Kramers doublet, which is given by $l^z\,{+}\,S^z\,{=}\,{\pm}\frac{1}{2}$, and the effective magnetic moment of Co$^{2+}$ is represented by ${\bm m}\,{=}\,g{\mu}_{\rm B}{\bm s}$ with the spin-$\frac{1}{2}$ operator $\bm s$~\cite{Abragam,Lines,Shiba}. In general, the $g$ factor is considerably anisotropic, and the total of the $g$ factors for the three different field directions is about 13~\cite{Abragam}, which is twice as large as that for conventional magnets. When the octahedral environment exhibits trigonal symmetry as in Ba$_3$CoSb$_2$O$_9$, the effective exchange interaction between fictitious spins ${\bm s}_i$ is described by the spin-$\frac{1}{2}$ $XXZ$ model~\cite{Lines,Shiba}
\begin{eqnarray}
{\cal H}_{\rm ex}=\sum_{<i,j>} \left[J_{\perp}\left\{s_i^xs_j^x+s_i^ys_j^y\right\}+J_{\parallel}s_i^zs_j^z\right].
\label{eq:int}
\end{eqnarray}
This interaction is Ising-like ($J_{\parallel}/J_{\perp}\,{>}\,1$) for ${\delta}/{\lambda}\,{<}\,0$, while it is $XY$-like ($J_{\parallel}/J_{\perp}\,{<}\,1$) for ${\delta}/{\lambda}\,{>}\,0$. The Heisenberg model ($J_{\parallel}/J_{\perp}\,{=}\,1$) is realized when ${\delta}\,{=}\,0$. We assume that the nearest-neighbor interaction on the triangular lattice is dominant in Ba$_3$CoSb$_2$O$_9$, as observed in isostructural Ba$_3$NiSb$_2$O$_9$~\cite{Shirata}. 

\begin{figure}[htbp]
\begin{center}
\vspace{-3.0 cm}
\includegraphics[scale =0.48]{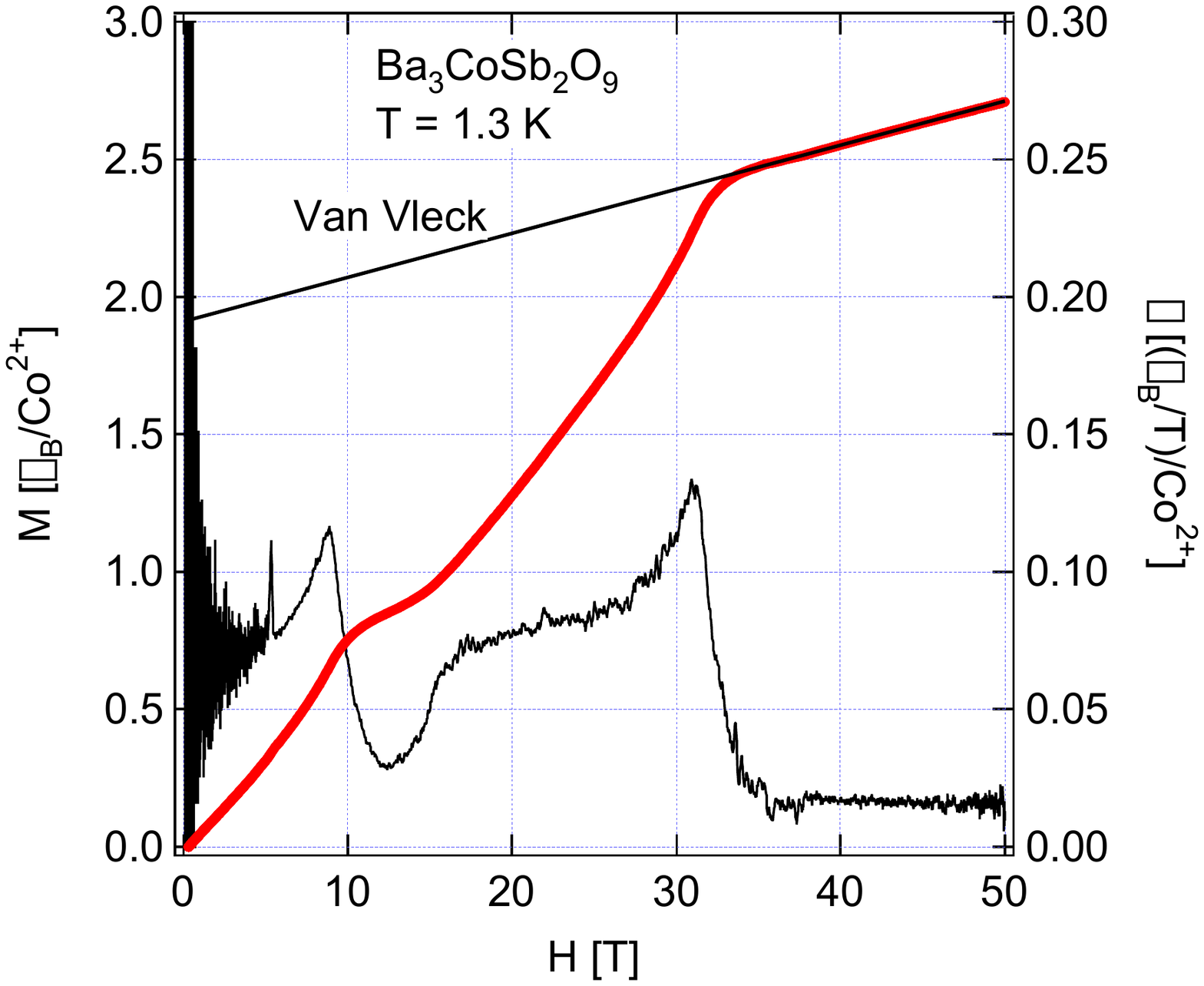}
\end{center}
\vspace{-4 cm}
\caption{(Color online) Raw magnetization curve of Ba$_3$CoSb$_2$O$_9$ powder measured at 1.3 K and derivative susceptibility $dM/dH$ vs magnetic field $H$. Dashed lines denote the Van Vleck paramagnetism evaluated from the magnetization slope above the saturation field $H_{\rm s}\,{=}\,31.9$ T. }
 \label{fig:MH1}
\end{figure}

\begin{figure}[htbp]
\begin{center}
\vspace{-3.0 cm}
\includegraphics[scale =0.48]{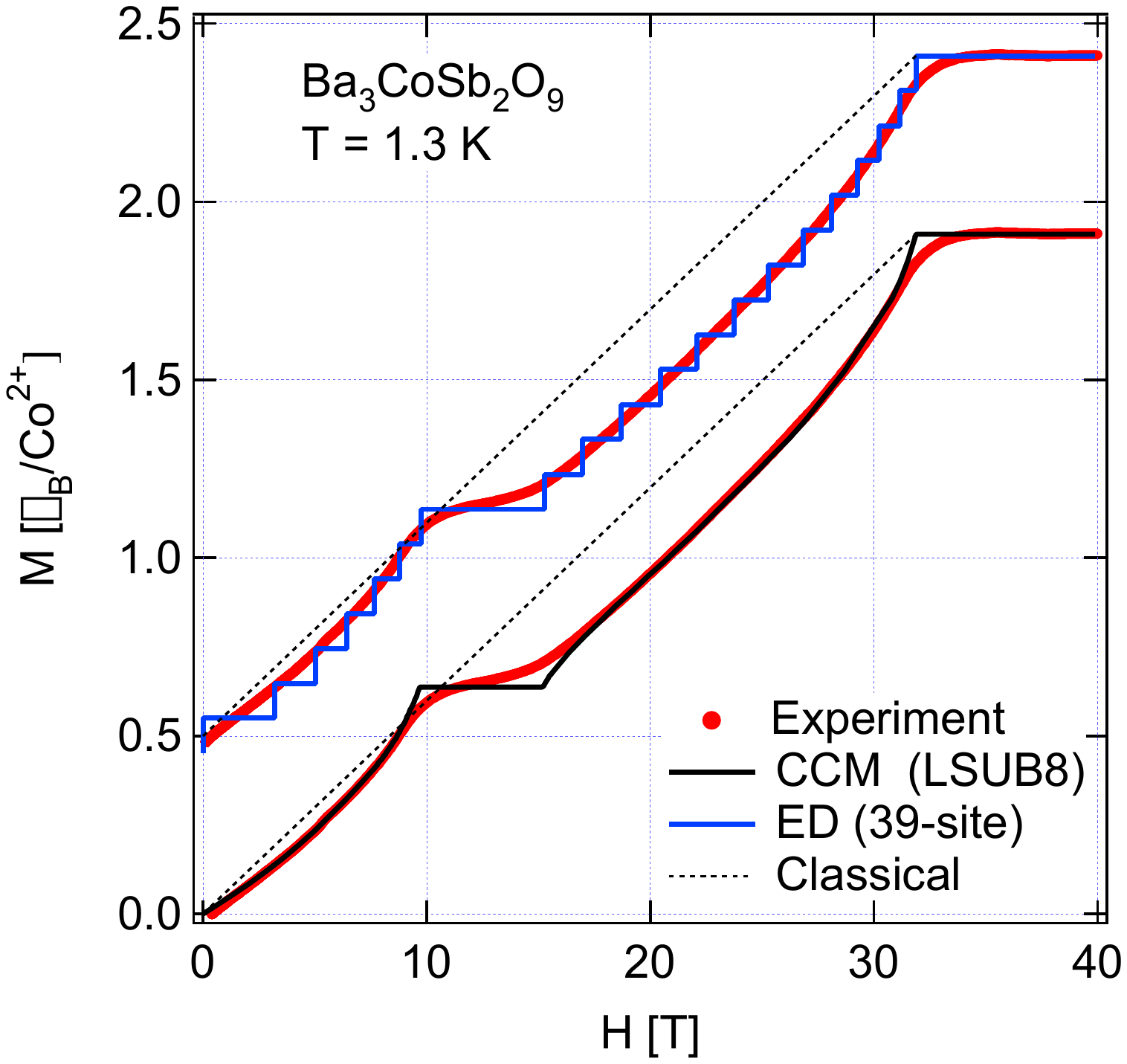}
\end{center}
\vspace{-4 cm}
\caption{(Color online) Magnetization curve corrected for Van Vleck paramagnetism. Thick black and blue lines denote the theoretical magnetization curves calculated by the higher order coupled cluster method (CCM)~\cite{Farnell} and exact diagonalization (ED) for a 39-site rhombic cluster~\cite{Sakai}, respectively. Thin dotted lines denote classical magnetization curves.}
 \label{fig:MH2}
\end{figure}

Ba$_3$CoSb$_2$O$_9$ powder was prepared via the chemical reaction 3BaCO$_3$\,{+}\,CoO\,{+}\,Sb$_2$O$_5$\,$\rightarrow$\,Ba$_3$CoSb$_2$O$_9$\,{+}\,3CO$_2$. Reagent-grade materials were mixed in stoichiometric quantities and calcined at 1100$^{\circ}$C for 20 h in air. Ba$_3$CoSb$_2$O$_9$ was sintered at 1200$^{\circ}$C for more than 20 h after being pressed into a pellet. To prepare single crystals, we packed the sintered Ba$_3$CoSb$_2$O$_9$ into a Pt tube of 9.6 mm inner diameter and 50 mm length. Small single crystals with dimensions of $1\,{\times}\,1\,{\times}\,1$ mm$^3$ were grown from the melt. The temperature of the furnace was lowered from 1700 to 1300$^{\circ}$C over two days. The samples obtained were examined by X-ray powder and single-crystal diffractions. 

The magnetic susceptibility of Ba$_3$CoSb$_2$O$_9$ powder was measured in the temperature range of $1.8\,{-}\,300$ K using a SQUID magnetometer (Quantum Design MPMS XL). High-field magnetization measurement in magnetic field of up to 53 T was performed at 4.2 and 1.3 K using an induction method with a multilayer pulse magnet at the Institute for Solid State Physics, University of Tokyo. The absolute value of the high-field magnetization was calibrated with the magnetization measured by the SQUID magnetometer. The specific heat of Ba$_3$CoSb$_2$O$_9$ single crystal was measured down to 0.4 K using a physical property measurement system (Quantum Design PPMS) by the relaxation method.

Figure \ref{fig:MH1} shows the raw magnetization curve and the derivative susceptibility for Ba$_3$CoSb$_2$O$_9$ powder measured at 1.3 K. The entire magnetization process was observed within a magnetic field of 53 T. The saturation of the Co$^{2+}$ spin occurs at $H_{\rm s}\,{=}\,31.9$ T. The increase in magnetization above $H_{\rm s}$ arises from the large temperature-independent Van Vleck paramagnetism characteristic of Co$^{2+}$ in the octahedral environment~\cite{Lines,Shiba}. From the magnetization slope above $H_{\rm s}$, the Van Vleck paramagnetic susceptibility was evaluated as ${\chi}_{\rm VV}\,{=}\,1.60\,{\times}\,10^{-2}\,{\rm ({\mu}_B/T)/Co^{2+}}\,{=}\,8.96\,{\times}\,10^{-3}$ emu/mol. The saturation magnetization was obtained as $M_{\rm s}\,{=}\,1.91\,{\rm {\mu}_B/Co^{2+}}$ by extrapolating the magnetization curve above $H_{\rm s}$ to a zero field (dashed line in Fig. \ref{fig:MH1}). 

Figure \ref{fig:MH2} shows the magnetization curves corrected for the Van Vleck paramagnetism. The quantum magnetization plateau is clearly observed at $\frac{1}{3}M_{\rm s}$. Thick dashed and solid lines denote fits by the higher order coupled cluster method (CCM)~\cite{Farnell} and exact diagonalization (ED) for a 39-site rhombic cluster~\cite{Sakai}, respectively. Both theories coincide with each other. The only adjustable parameters are saturation field Hs and saturation magnetization Ms. Although the experimental magnetization curve is smeared around the critical fields due to the finite temperature effect and the small anisotropies of the $g$ factor and the interaction, the agreement between the experimental and theoretical results is excellent. If the effective exchange interaction is strongly anisotropic, the magnetization process will strongly depend on the field direction, as observed in CsCoCl$_3$~\cite{Amaya}, and its spatial average will not agree with the theory for the 2D spin-$\frac{1}{2}$ TLHAF. The present result demonstrates that Ba$_3$CoSb$_2$O$_9$ closely approximates the 2D spin-$\frac{1}{2}$ TLHAF, although 3D magnetic ordering occurs at $T_{\rm N}\,{\simeq}\,3.8$ K owing to the small interlayer interaction. 

From the saturation magnetization and the relation $4.5J\,{=}\,g{\mu}_{\rm B}H_{\rm s}$, the average of the $g$ factor and the exchange constant were obtained as $g\,{=}\,3.82$ and $J/k_{\rm B}\,{=}\,18.2$ K, respectively. The magnetic field range of the $\frac{1}{3}$-magnetization plateau observed in Ba$_3$CoSb$_2$O$_9$ agrees with that of $0.306\,{<}\,H/H_{\rm s}\,{<}\,0.479$ predicted by the CCM~\cite{Farnell} and ED~\cite{Honecker,Sakai}. This field range is much larger than that of $0.45\,{<}\,H/H_{\rm s}\,{<}\,0.50$ observed in Cs$_2$CuBr$_4$ with a spatially anisotropic triangular lattice ($J_2/J_1\,{=}\,0.74$)~\cite{Ono1,Fortune} and is twice as large as that of $0.317\,{<}\,H/H_{\rm s}\,{<}\,0.413$ for the spin-1 case~\cite{Shirata}.

\begin{figure}[htbp]
\begin{center}
\vspace{-3.0 cm}
\includegraphics[scale =0.48]{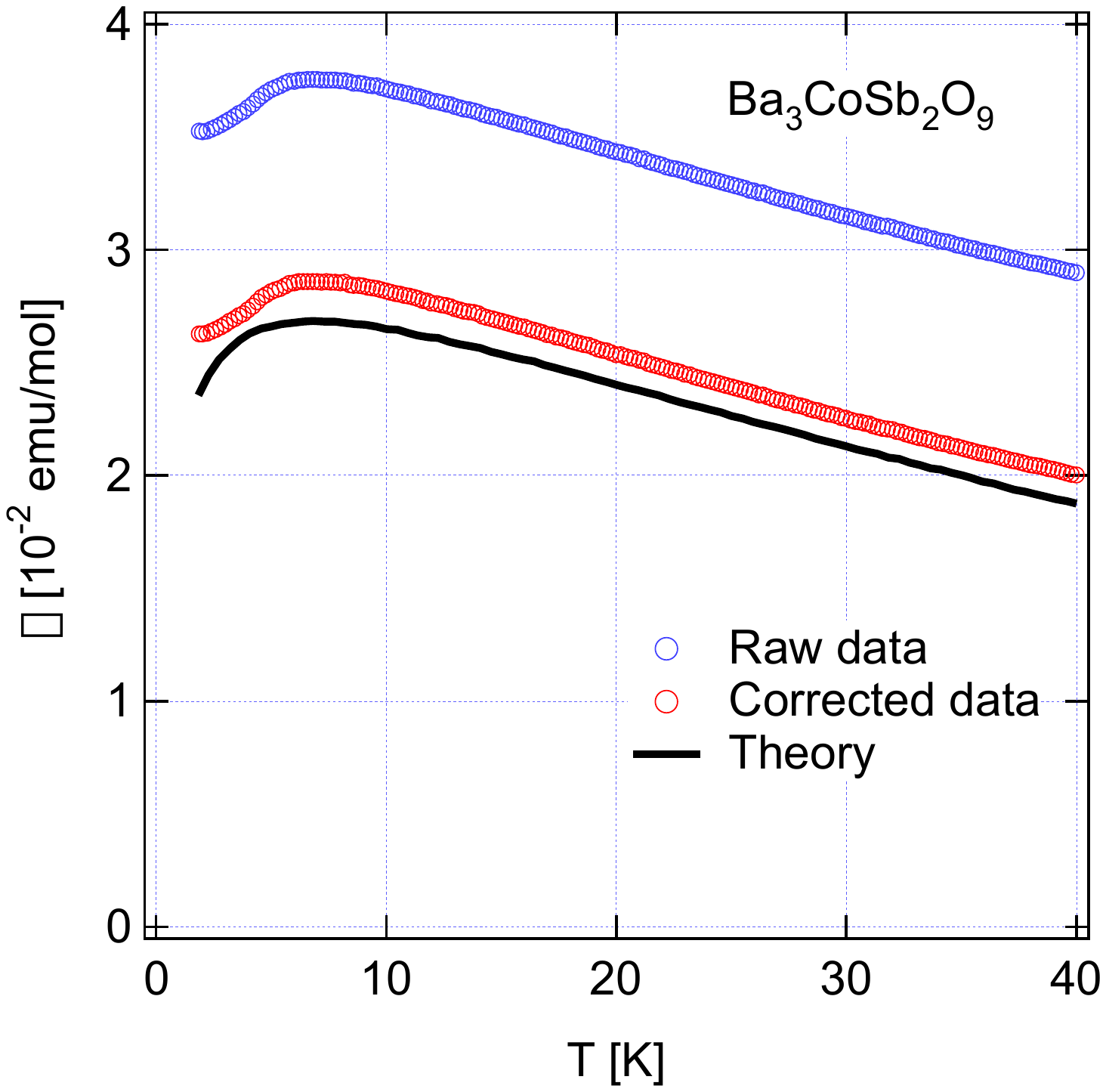}
\end{center}
\vspace{-4 cm}
\caption{(Color online) Temperature dependence of magnetic susceptibilities for Ba$_3$CoSb$_2$O$_9$ obtained before and after correction of Van Vleck paramagnetism. The solid line denotes the theoretical susceptibility calculated by series expansion~\cite{Elstner} with $J/k_{\rm B}=18.2$ K and $g=3.82$.}
 \label{fig:sus}
\end{figure}

Figure \ref{fig:sus} shows the magnetic susceptibilities of Ba$_3$CoSb$_2$O$_9$ powder obtained before and after the correction of the Van Vleck paramagnetic susceptibility of ${\chi}_{\rm VV}\,{=}\,8.96\,{\times}\,10^{-3}$ emu/mol. We plotted the susceptibility data for $T\,{\leq}\,40$ K, where the spin-$\frac{1}{2}$ description of the magnetic moment is valid. The contribution of the Van Vleck paramagnetic susceptibility is $\frac{1}{4}\,{-}\,\frac{1}{3}$ of the raw magnetic susceptibility, and thus, its correction is essential for evaluating the intrinsic magnetic susceptibility. The magnetic susceptibility has a rounded maximum at 7 K, characteristic of a low-dimensional antiferromagnet. The solid line in Fig. \ref{fig:sus} indicates the theoretical susceptibility of the 2D spin-$\frac{1}{2}$ TLHAF calculated by series expansion~\cite{Elstner} with $J/k_{\rm B}\,{=}\,18.2$ K and $g\,{=}\,3.82$, which were obtained from the present high-field magnetization measurements. The experimental and theoretical magnetic susceptibilities are consistent, although the theoretical susceptibility is smaller than the experimental susceptibility.

Using a small single crystal, we also measured the specific heat to investigate the nature of the magnetic ordering. Figure 3b shows the low-temperature specific heat measured at a zero magnetic field. Sharp peaks indicative of magnetic phase transitions were observed at around 3.8 K. As shown in the inset, Ba$_3$CoSb$_2$O$_9$ undergoes three magnetic phase transitions at $T_{\rm N1}\,{=}\,3.82, T_{\rm N2}\,{=}\,3.79$ and $T_{\rm N3}\,{=}\,3.71$ K. In the Heisenberg-like triangular-lattice antiferromagnet, successive phase transitions occur when the magnetic anisotropy is of the easy-axis type, while a single transition arises for easy-plane anisotropy~\cite{Matsubara,Miyashita}. The successive phase transitions observed show the presence of easy-axis anisotropy, which is consistent with the 120$^{\circ}$ spin structure in a plane including the $c$ axis observed by Doi {\it et al.}~\cite{Doi}. However, Ba$_3$CoSb$_2$O$_9$ differs from other triangular-lattice antiferromagnet with easy-axis anisotropy in ordering process. Usually, the magnetic ordering occurs in two steps as observed in CsNiCl$_3$~\cite{Adachi,Beckmann}, while in Ba$_3$CoSb$_2$O$_9$, it occurs in three steps.

The reduced temperature range of the intermediate phase $(T_{\rm N1}\,{-}\,T_{\rm N3})/T_{\rm N1}$ is determined from the ratio of the anisotropic term $(J_{\parallel}\,{-}\,J_{\perp})$ to the isotropic term $(J_{\perp})$ in the exchange interaction~\cite{Matsubara,Miyashita}. The very narrow temperature range of the intermediate phase in Ba$_3$CoSb$_2$O$_9$ means that the anisotropic term is much smaller than the isotropic term. The effective exchange interaction of eq.\,(\ref{eq:int}) is strongly anisotropic in typical cobalt substances~\cite{Lines,Shiba}, but in Ba$_3$CoSb$_2$O$_9$, it is accidentally close to the Heisenberg model. 

\begin{figure}[htbp]
\begin{center}
\vspace{-3.0 cm}
\includegraphics[scale =0.48]{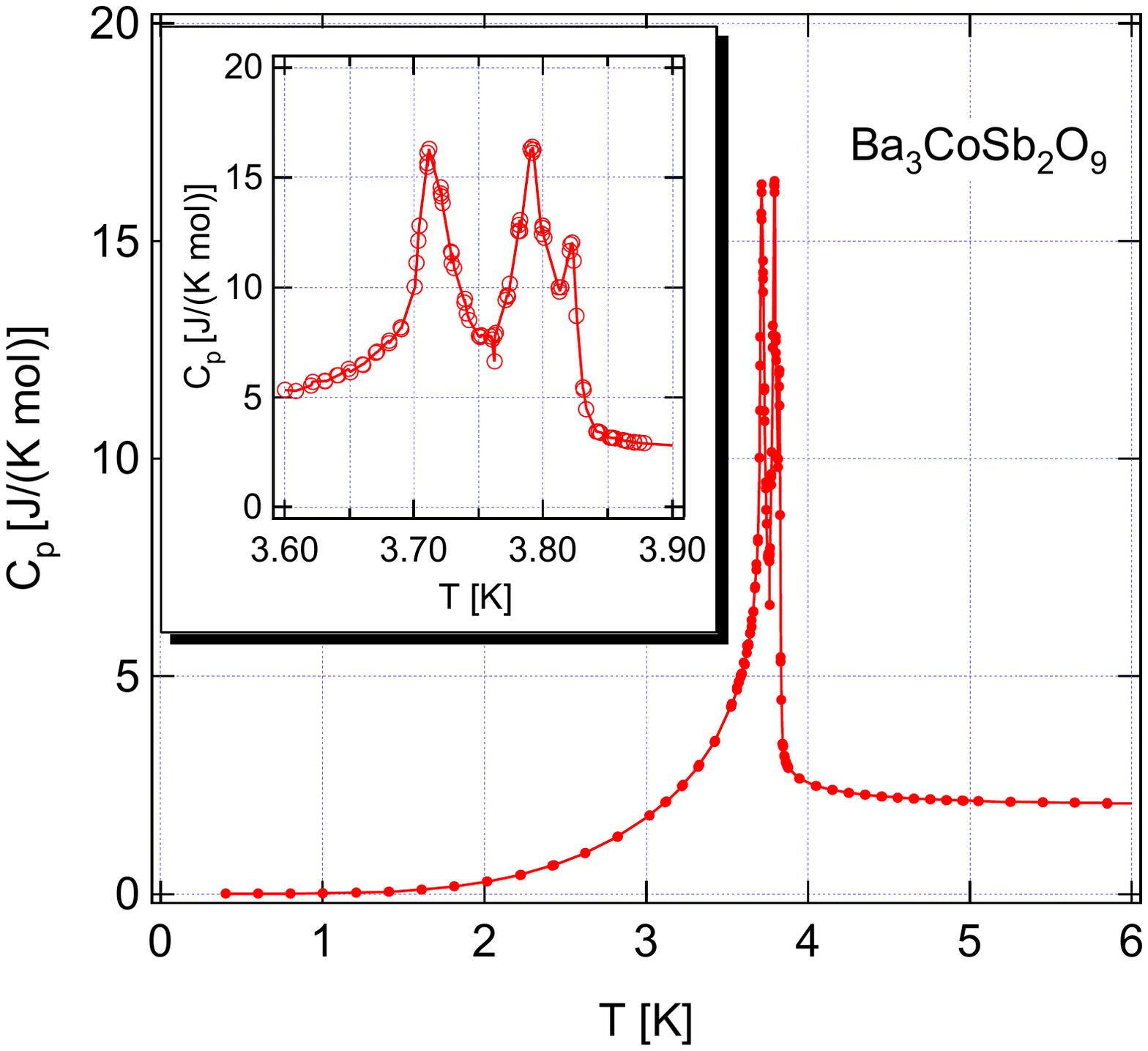}
\end{center}
\vspace{-4 cm}
\caption{(Color online) Low-temperature specific heat of Ba$_3$CoSb$_2$O$_9$ measured at zero magnetic field. The inset shows an expansion of the graph around 3.8 K.}
 \label{fig:heat}
\end{figure}

In conclusion, we have shown that the entire magnetization process and the temperature dependence of  the magnetic susceptibility for Ba$_3$CoSb$_2$O$_9$ agree well with theoretical results for spin-$\frac{1}{2}$ TLHAF, and that Ba$_3$CoSb$_2$O$_9$ undergoes three magnetic phase transitions with very narrow intermediate phases. These results demonstrate that the spin-$\frac{1}{2}$ TLHAF is actually realized in Ba$_3$CoSb$_2$O$_9$. Conversely, this work verifies recent theory on the magnetization process for spin-$\frac{1}{2}$ TLHAF. Therefore, Ba$_3$CoSb$_2$O$_9$ is expected to be useful for verifying quantum-fluctuation-assisted spin states in magnetic fields~\cite{Chubukov,Nikuni,Alicea} and for exploring new quantum aspects of the spin-$\frac{1}{2}$ TLHAF, such as negative quantum renormalization and the singularity of magnetic excitations~\cite{Starykh,Zheng,Chernyshev}.

We express our sincere thanks to D. J. J. Farnell, R. Zinke, J. Schulenburg, J. Richter, H. Nakano and T. Sakai for showing us their theoretical calculations of the magnetization process and to Y. Takano and I. Umegaki for useful discussions. We are deeply indebted to T. Shimizu and M. Itoh for their support of single-crystal X-ray diffraction. This work was supported by a Grant-in-Aid for Scientific Research from the Japan Society for the Promotion of Science, and the Global COE Program gNanoscience and Quantum Physicsh at TIT funded by the Ministry of Education, Culture, Sports, Science and Technology of Japan. H.T. was supported by a grant from the Mitsubishi Foundation.

\end{document}